# Energy-Aware Routing for E-Textile Applications[†]

Jung-Chun Kao and Radu Marculescu

*Carnegie Mellon University*
*Pittsburgh, PA 15213, U.S.A.*
{*jungchuk, radum*}*@ece.cmu.edu*

## Abstract

*As the scale of electronic devices shrinks, "electronic textiles" (e-textiles) will make possible a wide variety of novel applications which are currently unfeasible. Due to the wearability concerns, low-power techniques are critical for e-textile applications. In this paper, we address the issue of the energy-aware routing for e-textile platforms and propose an efficient algorithm to solve it. The platform we consider consists of dedicated components for e-textiles, including computational modules, dedicated transmission lines and thin-film batteries on fiber substrates. Furthermore, we derive an analytical upper bound for the achievable number of jobs completed over all possible routing strategies. From a practical standpoint, for the Advanced Encryption Standard (AES) cipher, the routing technique we propose achieves about fifty percent of this analytical upper bound. Moreover, compared to the non-energy-aware counterpart, our routing technique increases the number of encryption jobs completed by one order of magnitude.*

## 1. Introduction

The scaling of device technologies has opened research opportunities in the area of pervasive computing [4]. In particular, it became possible to fabricate flexible materials with sophisticated computing and communication devices embedded into them. Such computational fabrics, called e-textiles, are expected to have many applications ranging from consumer electronics, to military, health and security devices.

For e-textiles, the system has to be not only light and mobile, but also withstand wear-and-tear due to frequent washing. Due to the potential stress upon e-textile's interconnects, one solution is to use a communication network instead of the traditional bus-based architecture. Besides reliable operation, e-textiles have strict processing, storage, energy and size constraints per computational node. Thus, in order to be able to handle complicated applications, electronic textiles need to rely on low-power techniques. The target application that runs on an e-textile platform must be appropriately partitioned and able to self-manage as a whole. These additional requirements lead to an interesting and unique routing problem which is the main objective of this paper.

Advanced Encryption Standard (AES) [7] is selected as the driver application running onto an e-textile platform. This selection is motivated by two reasons: First, AES is a robust encryption algorithm so since data secrecy plays a critical role in pervasive computing, exploring the potential of AES for distributed implementations on e-textile fabrics becomes very important. Second, in June 2004, IEEE ratified the wireless local area network (WLAN) standard, 802.11i which requires AES for data encryption. This further increases the potential of AES for future pervasive applications.

Our approach to study the distributed implementation of AES on e-textiles consists on several steps: First, we divide the AES cipher into several modules (each performing an unique function) that we design and simulate in Verilog. Second, we develop an energy-aware routing strategy. This consists of a mapping technique from application modules to architecture nodes in a mesh-based topology of various sizes, a control mechanism based on a TDMA scheme, and the *online* general-purpose energy-aware routing (*EAR*) algorithm. A cycle-accurate network simulator, *et_sim*, has been developed as a by-product for e-textile platforms. The electrical characteristics of dedicated components extracted from our designs, plus data from [6,10] are fed into *et_sim* to compare EAR results with the non-energy-aware routing counterpart.

From a theoretical standpoint, our contribution is twofold: First, we formulate the routing problem for e-textile applications and develop an online energy-aware routing algorithm to solve it. Second, we derive an analytical upper bound for the achievable number of jobs completed in arbitrary architectures. According to the simulation results, EAR achieves approximately 50% of the analytical upper bound and exceeds its non-energy-aware counterpart by a factor between 5 to 15 times. We point out that the methodology and theoretical results presented here apply to any e-textile distributed system, either wearable or non-wearable (e.g. e-textiles hanging on the walls, etc).

The remainder of the paper is organized as follows. In Section 2, related work is summarized. We formulate the routing problem in Section 3. In Section 4, we present the analytical upper bound for all possible routing strategies. Section 5 outlines the architectural issues, the battery model and detailed platform description. The general-purpose energy-aware routing algorithm is discussed in Section 6. We present the experimental results for the AES cipher and compare them with the analytical upper bound in Section 7. In Section 8, we conclude by summarizing our main contribution.

## 2. Related Work

There has been some recent work on embedding complicated electrical components into wearable fabrics. In [1,2], the authors demonstrate the idea of attaching off-the-shelf electrical components to traditional clothing materials and also provide the method by which user interfaces (and even chip packages) may be woven directly into fabric during textile manufacturing.

[†] This research was supported in part by MARCO/DARPA Gigascale Systems Research Center (GSRC) under grant no. 2003-DT-660.

On the other hand, devices dedicated for e-textiles such as textile transmission lines (e.g. [6]) and batteries (e.g. thin-film batteries [10,11]) are currently under development. The routing of electrical power and communication through a wearable fabric is addressed in [3]. It provides a detailed account of the physical and electrical components for routing electricity through suspenders made of fabric and embedded conductive strands, as well as a data-link layer protocol on a Controller Area Network (CAN) bus.

Several architectures with distributed deployment of batteries are proposed in [5]. Adaptive techniques, such as code migration and remote execution, applied to redundantly deployed nodes have also been proposed to increase the operational lifetime of the e-textile applications.

In contrast to these previous efforts, we investigate techniques for energy-aware routing to increase the operational lifetime of a system and the achievable number of completed jobs. Several energy-aware routing algorithms have been proposed for wireless ad-hoc/sensor networks in [12,13]. However, they *do not* apply to e-textile platforms because of the inherent difference in the cost functions that characterize wireless and wired environments, and because of their high requirement of computational and memory resources. We follow the same general distributed deployment scheme of batteries as in [5], but due to potential link failures, the target architecture in our case is a network instead of a bus-based architecture. Also, unlike [5], fixed and re-programmable devices are supported as computational nodes in our architecture. These differences make our routing problem quite unique.

## 3. Problem Description

We assume that *i)* the target application is partitioned into several modules which have been customized or are available as IP cores in a library, and *ii)* the target e-textile system is composed of a *wired* communication network which connects many active and idling nodes. Each module performs an unique (fixed) function and cooperates with other modules to complete the job by exchanging packets of fixed length. Each node is an instance of exactly one module; it is possible to have, however, duplicates of modules across the network. The total number of nodes that implement different modules in the network cannot exceed the total number of available nodes.

It is assumed that each node has its own attached battery and, for simplicity, all batteries have the same initial capacity. An *operation* is defined as the act of computation at any module and the subsequent act of communication until the originated packet arrives at the next node which can be either an intermediate node or the destination node. Typically, each module needs to perform several operations before a job is completed. A node is considered dead when its attached battery is completely depleted. The target system dies when the critical nodes become dead.

In general, the hardware modules can be either fixed or reprogrammable, but the remapping techniques (e.g. code migration and remote execution [5]) on reprogrammable modules are not considered in this paper. Instead, it is assumed that, once determined, the application mapping onto architecture remains fixed. This avoids not only the difficult prediction of the overhead caused by the remapping techniques (which is highly dependent on architecture and application itself), but also the requirement of using reprogrammable components. Our main objective is to design an energy-aware routing strategy that can increase the number of jobs completed for e-textile applications. Finding the appropriate routing strategy involves several design issues such as choosing the appropriate network topology, control mechanism, mapping technique, etc. To better present the formulation of the routing strategy problem, we summarize the basic parameters in Table 1.

**Table 1: Parameter notation**

| Parameter | Description |
|---|---|
| $RS$ | Routing strategy which consists of network topology, mapping technique, control mechanism and routing algorithm |
| $J^{(RS)}$ | The number of jobs completed under the routing strategy $RS$ |
| $B$ | Initial capacity of each battery (battery budget) |
| $K$ | The maximum number of available nodes (node budget) |
| $p$ | The number of (distinct) modules |
| $f_i$ | The number of operations that module $i$ has to run before a job is completed where $1 \leq i \leq p$ |
| $E_i$ | Energy consumption of module $i$ per act of computation where $1 \leq i \leq p$ |
| $S_i$ | The set of nodes at which duplicates of module $i$ are located where $1 \leq i \leq p$ |
| $n_i$ | The number of duplicates of module $i$ in the target topology, i.e. There are $n_i$ nodes mapped to module $i$ where $1 \leq i \leq p$ |
| $c_i$ | Energy consumption per act of communication originated from module $i$ where $1 \leq i \leq p$ |
| $C_j$ | Total Energy consumption at node $j$ during communication (either transmitting the packet originated from node $j$ or relaying packets originated from other nodes) before the target system dies, where $1 \leq j \leq K$ |
| $OH_j$ | The induced overhead in terms of energy consumption by the routing strategy $RS$ at node $j$ where $1 \leq j \leq K$ |
| $\varepsilon_i$ | The normalized energy consumption of module $i$, which is defined as the total energy that module $i$ (more precisely, all nodes mapped to module $i$) consumes before a job is completed |

Some parameters in Table 1 depend only on application and architecture. For instance, an encryption round (job) for the 128-bit AES (see Fig 1) takes 10 operations of type *Subbytes/Shiftrows*, 9 operations of type *Mixcolumns* and 11 operations of type *Addroundkey*. Therefore $f_1$, $f_2$ and $f_3$ are 10, 9 and 11 respectively. Other parameters, however, depend also on the routing algorithm and control mechanism. The induced overhead, $OH_j$, and the total energy consumption during communication, $C_j$, are such examples.

With these notations, the problem of determining the routing strategy for performance maximization (in terms of number of completed jobs), under resource constraints, can be formulated as follows:
**Given:**
 Data flow of the target application, specifically $p$ and $f_i$'s
 Modules energy consumption per act of computation, $E_i$'s
 The battery budget, $B$, and the node budget, $K$

**Determine:**

The optimal routing strategy which maximizes the number of jobs completed over *all possible* routing strategies; that is,

$$\max_{RS}(J^{(RS)})$$

such that

$$\sum_{i=1}^{p} n_i \leq K$$
$$\sum_{j \in S_i} x_j / f_i \geq J^{(RS)}, \forall i = 1, 2, \ldots, p$$
$$B \geq E_{i(j)} x_j + C_j + OH_j$$

where $x_j$ is the actual number of operations at node $j$ before the target system dies and the subscript $i(j)$ denotes the type of module that node $j$ implements.

The first condition in the above formulation requires that the total number of nodes is less or equal to the maximum number of available nodes. The second condition ensures that $J^{(RS)}$ jobs are completed under the routing strategy $RS$. The last condition guarantees that the target system remains alive before $J^{(RS)}$ jobs are completed. Among all possible routing strategies, the one which satisfies these three conditions and maximizes the number of completed jobs represents the optimal solution.

## 4. Analytical Results on Routing Strategy

In this section, we derive an analytic upper bound on the achievable number of jobs completed over all routing strategies. To this end, we first construct the *ideal* routing strategy $RS^*$ which has the following features:

  i. The topology of $RS^*$ is chosen to match the data flow of the target application.
  ii. The mapping from modules to nodes is considered optimal. Specifically, for each $i$, the number of duplicates of module $i$, $n_i$, is optimal. Furthermore, for each $i$, we expand the domain of $n_i$ from positive integers to positive real numbers.
  iii. An incomplete operation due to node battery depletion can, without incurring any cost, continue the remaining fraction of operation at a living node that performs the same functionality.
  iv. There is no overhead of the ideal routing strategy caused by the control mechanism.

Feature *i* implies that $RS^*$ consumes the least amount of energy during communication. Feature *ii* ensures that the more power a module consumes, the higher proportion of nodes will be mapped to that module. Feature *iii* implies that the target system is alive as long as, for each module, there exists (at least) a live node mapped to that module.

Features *i*, *iii* and *iv* reduce the complex routing strategy problem to the problem of optimizing the number of nodes for each type of module. To complete a job, module $i$ has to perform $f_i$ operations, each consuming $E_i + c_i$ on computation and communication, so the energy consumption sums up to the normalized energy consumption, $\varepsilon_i = f_i (E_i + c_i)$. Since the target system under $RS^*$ dies when all nodes where some application module is mapped to become dead, the achievable number of completed jobs under an arbitrary routing strategy $RS$ is

$$J^{(RS)} \leq J^{(RS^*)} = \left\lfloor \max_{\underline{n} \in (R^+)^p : \sum_{i=1}^{p} n_i \leq K} \min(\frac{n_1 B}{\varepsilon_1}, \frac{n_2 B}{\varepsilon_2}, \cdots, \frac{n_p B}{\varepsilon_p}) \right\rfloor$$

$$\leq \max_{\underline{n} \in (R^+)^p : \sum_{i=1}^{p} n_i \leq K} \min(\frac{n_1 B}{\varepsilon_1}, \frac{n_2 B}{\varepsilon_2}, \cdots, \frac{n_p B}{\varepsilon_p}) \quad (1)$$

where $\underline{n}$ is a vector defined as $(n_1, n_2, \ldots, n_p)$.

Due to limited space, we state Theorem 1 without proof.

**Theorem 1**: (Upper bound for the achievable number of completed jobs) Given the application parameters $p$ and $f_i$'s, the energy consumption for each module $E_i$, the battery budget $B$ and the node budget $K$, the maximum number of completed jobs is given by

$$J^* = \frac{K B}{\sum_{i=1}^{p} \varepsilon_i} \quad (2)$$

where $\varepsilon_i$ is the normalized energy consumption of module $i$. Furthermore, the optimal number of duplicates of modules $i$ is given by

$$n_i^* = \frac{K \varepsilon_i}{\sum_{j=1}^{p} \varepsilon_j} \quad (3)$$

Theorem 1 not only gives a tight upper bound for the achievable number of completed jobs, but also reveals an important design rule: for each $i$, the optimal number $n_i^*$ of duplicates of module $i$ is proportional to the corresponding value of the normalized energy consumption, $\varepsilon_i$.

## 5. Platform Description

Although our energy-aware routing strategy can be applied to any application, we focus on the AES cipher as an illustrative example of our work. For a fair comparison, the proposed energy-aware routing strategy and its non-energy-aware counterpart are kept exactly the same except their routing algorithms which are the energy-aware routing algorithm (EAR) and the shortest-distance routing algorithm (SDR), respectively.

### 5.1 Energy Modeling

E-textile systems have to be tiny, light, flexible and slim so that they can be easily embedded into the fabric. Due to the above reasons, the novel thin-film battery [10,11] is a good candidate for the attached battery on e-textile platforms. The battery modeling for the thin-film battery is addressed in this section. We also present the energy models of the computational modules and the transmission lines for the AES cipher on e-textiles. Using measured data, we find out that, for the AES cipher, the power consumed on the transmission lines is *not* negligible compared with the power consumed in the computational modules. This suggests that, for e-textiles, the remaining battery capacity, as well as the distances of routing paths should be considered when making the routing decisions.

#### 5.1.1 AES Partitioning and Computation Energy Consumption

We outline the pseudo code of AES cipher in Fig 1. For 128-bit AES, $Nb$=4 and $Nr$=10. Because of limited capability for raw processing and battery capacity, the whole system has to be partitioned into several modules. None of these modules should consume a large amount of power. Finally, the following partitioning scheme is used.
- Module 1: *SubBytes*( ) / *ShiftRows*( )
- Module 2: *MixColumns*( )
- Module 3: *KeyExpansion* / *AddRoundKey*( )

We specified all these modules in Verilog and synthesized them with the Synopsys Design Compiler using a 0.16μm technology. While these modules can operate at clock frequencies up to 233MHz, the power consumption at 100MHz was measured. The energy consumption values, per act of computation, are $E_1$=120.1pJ, $E_2$=73.34pJ, $E_3$=176.55pJ for modules 1, 2 and 3, respectively.

```
Cipher(byte in[4*Nb], byte out[4*Nb], word w[Nb*(Nr+1)])
begin
    byte state[4,Nb]
    state = in
    AddRoundKey(state, w[0, Nb-1])

    for round = 1 step 1 to Nr–1
        SubBytes(state)
        ShiftRows(state)
        MixColumns(state)
        AddRoundKey(state, w[round*Nb, (round+1)*Nb-1])
    end for

    SubBytes(state)
    ShiftRows(state)
    AddRoundKey(state, w[Nr*Nb, (Nr+1)*Nb-1])
    out = state
end
```
**Fig 1: The pseudo code of the AES cipher**

### 5.1.2 Electrical Characteristics of Transmission Lines and Communication Energy Consumption

The electrical characteristics of dedicated textile transmission lines of various lengths are extracted from [6]. The fabrics used contain polyester yarns that are twisted with one copper thread of 40μm diameter and are insulated with a polyesterimide coating. We performed SPICE simulations and found out that the energy consumption values, per bit-switching activity, are 0.4472pJ, 4.4472pJ, 11.867pJ and 53.082pJ for transmission lines of 1cm, 10cm, 20cm and 100cm, respectively. These values multiplied by the packet size represent the energy consumed on transmitting a packet over these transmission lines and are fed into our network simulator, *et_sim*.

### 5.1.3 Battery Modeling

Fig 2 shows the discharge voltage profile of a Li-free thin-film battery [10]. The discharging characteristic together with the discrete-time model in [8] which closely approximates the behavior of its circuit-level continuous-time counterpart is implemented in our network simulator, *et_sim*. This way, the accuracy is acceptable (within 15%), since the actual capacity of any group cells may vary as much as 20% even between identical units [8].

To reduce the simulation time, the initial (nominal) capacity of the thin-film battery is reduced to 60000pJ. We also assume that the corresponding discharging voltage profile shrinks proportionally in the horizontal direction. We assume that a node is dead after the output voltage of its attached battery drops below 3.0 Volts and the remaining energy stored in the attached battery is wasted.

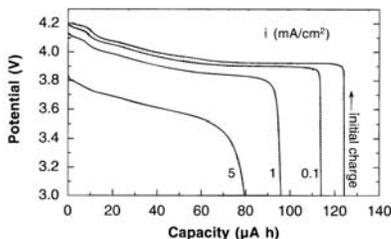

**Fig 2: Discharge curve of thin-film batteries (from [10])**

## 5.2 Mapping the AES Application onto the Mesh Architecture

As Theorem 1 shows, the normalized energy consumption per module ($\varepsilon_i$) is a key factor for the mapping from application modules to network nodes. Based on Theorem 1, a large number of nodes are mapped to module 3 which consumes the highest normalized energy among all three application modules.

Assuming any node with coordinates ($x$, $y$), our mapping strategy is to map that node to module 1 if $m(x)+m(y)=2$, to module 2 if $m(x)+m(y)=0$, and to module 3 if $m(x)+m(y)=1$ where $m(x)$ is defined as $x\ modulo\ 2$. Fig 3(a) sketches a smart shirt with several blocks connected through a wired network; Fig 3(b) zooms in to the region where the AES modules are mapped using a 4x4 mesh network.

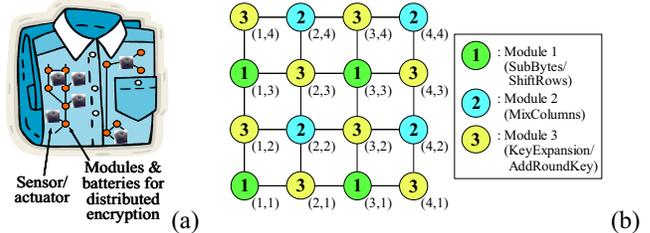

**Fig 3: (a) Network-based architectures, (b) The mapping of AES using a 4-by-4 mesh network**

## 5.3 Control Mechanism

A centralized control mechanism based on a time division multiple access (TDMA) scheme (see Fig 4) has been implemented. It consists of several active and idle centralized controllers, many AES nodes and a shared communication medium. The AES nodes periodically report their respective statuses to the central controllers during their own upload slots. The active central controllers figure out the routing paths for all nodes based on the reported information. After the routing paths are determined, the information about the next hops on routing paths is sent to corresponding regular nodes through the shared medium in the coming downloading phase.

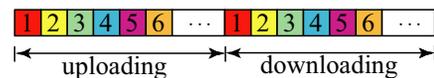

**Fig 4: TDMA scheme used to implement the control mechanism**

The shared communication medium is used only for exchanging the control information. Under the TDMA scheme, the width of shared medium can be very narrow, for instance, only 2-bit wide. In addition, compared with the significant performance gain (see Fig 7 in Sec 7), the percentage of energy consumed on the shared medium is small. Therefore, backing up the information sent over the shared medium, in case of potential failures, is feasible even under the strict resource constraints that characterize e-textile systems.

We include a deadlock recovery mechanism in the TDMA scheme: when a job stays at a node for more than a threshold period, that node needs to report the occurrence of deadlock during its next upload slot. The central controller sends then the new routing instruction to that node to redirect the job along an unlocked path during the next download phase.

## 6. EAR and SDR Algorithms

The EAR algorithm and its shortest distance routing counterpart (SDR) have been developed to decide the routing paths *on-line*, based on the reported system information. Both of them have the capability of recovery from deadlock.

When the currently reported system information differs from the previous one, the central controller executes the routing algorithm in order to instruct the nodes on how to modify their routing tables. SDR generally selects the shortest path while EAR determines the path based on the reported battery information as well. Both of SDR and EAR are composed of three phases that will be detailed next.

In the first phase, for both SDR and EAR, a *weight* is assigned to each directed interconnect. The network topology is represented as a directed graph $G = (V, E)$ where $V$ is the set of vertices and $E$ is the set of edges. For convenience, our algorithms use an adjacency-matrix representation. The edge weight matrix, $W^{(SDR)} = [W_{ij}^{(SDR)}]$, of SDR is set to

$$W_{ij}^{(SDR)} = \begin{cases} 0 & \text{if } i = j \\ L_{ij} & \text{if } i \neq j \text{ and } (i,j) \in E \\ \infty & \text{if } i \neq j \text{ and } (i,j) \notin E \end{cases}$$

where $L_{ij}$ is the length of the directed interconnect. One the other hand, the edge weight matrix $W^{(EAR)} = [W_{ij}^{(EAR)}]$ of EAR is set to

$$W_{ij}^{(EAR)} = \begin{cases} 0 & \text{if } i = j \\ f(N_B(j)) L_{ij} & \text{if } i \neq j \text{ and } (i,j) \in E \\ \infty & \text{if } i \neq j \text{ and } (i,j) \notin E \end{cases}$$

where $N_B(j) \in Z^+$ is the reported battery level of node $j$, $0 \leq N_B(j) < N_B$, and $f(\cdot)$ is a weighting function. In our experiments, $f(n)$ is set to $2^{Q(N_B - 1 - n)}$ where $Q > 0$ is a constant to strengthen the impact of the battery information.

```
1  D^(0) = W
2  for n = 1 to K
3    for i = 1 to K
4      for j = 1 to K
5        D_ij^(n) = min(D_ij^(n-1), D_in^(n-1) + D_nj^(n-1))
6        S_ij^(n) = { S_ij^(n-1)  if D_ij^(n-1) ≤ D_in^(n-1) + D_nj^(n-1)
                     S_in^(n-1)  if D_ij^(n-1) > D_in^(n-1) + D_nj^(n-1)
7  return D^(K) and S^(K)
```

**Fig 5: The 2nd phase of EAR and SDR algorithms**

During the second phase, for all pairs of nodes, we compute the $K$-by-$K$ shortest-path matrix $D = [D_{ij}]$ and the $K$-by-$K$ successor matrix $S = S_{ij}$, for $1 \leq i, j \leq K$, where $K$ denotes the budget of nodes, $D_{ij}$ is the shortest distance from node $i$ to node $j$, and $S_{ij}$ is the successor of node $i$ along the shortest path to node $j$. A variation of the Floyd-Warshall algorithm of complexity $O(n^3)$ (see [9]) is used to compute all-pairs shortest paths and their successors; the pseudo code is given in Fig 5. In the pseudo code in Fig 5, $W$ is set to $W^{(SDR)}$ for SDR and is set to $W^{(EAR)}$ for EAR.

The third phase has to determine the destination node among the nodes which have the same functionality and avoid using the ports which are currently in a deadlock state. The pseudo code is shown in Fig 6.

The recovery from deadlock is ensured because of the *if-branch* in line 5 of the pseudo code. The third phase yields a running time of $O(n^2)$. This is because the total number of executions of line 5 is $K(n_1+n_2+...+n_p) = K^2$. The hidden constant in the third phase is close to zero when few deadlock and/or congestion situations occur (see lines 5-8 in Fig 6). Even in case of severe deadlock/congestion, the hidden constant in the third phase is at most equal to that in the second phase.

```
RT_j(i) ≡ the successor of node j on a shortest path to
         S_i in the routing table at node j
1  for n = 1 to K
2    for i = 1 to p
3      dist = ∞
4      for j ∈ S_i
5        if node n is not in deadlock or S_nj ≠ RT_n(i)
6          if dist > D_nj
7            suc = S_nj
8            dist = D_nj
9      RT_n(i) = suc
```

**Fig 6: The 3rd phase of EAR and SDR algorithms**

From the above arguments, we know that, for either EAR or SDR, the complexity is $O(n^3)$, the hidden constants are small and most of the running time is spent in the second phase. Thus, EAR and SDR are practical for graphs consisting of tens to a few hundreds of nodes.

## 7. Experimental Results

To evaluate the performance gain that can be achieved under EAR, we simulate the AES cipher using mesh networks of various sizes. The results are compared with those obtained by using its non-energy-aware counterpart, SDR. The results are also compared with the analytical upper bound to illustrate how much room for further improvement still exists. In addition, multiple concurrent jobs are fed into the target system to see the effectiveness of the developed deadlock recovery mechanism.

A cycle-accurate network simulator, *et_sim*, was implemented. *et_sim* supports, in default mode, any 2D mesh network with the mapping technique described in Sec 5.2. Since many factors (e.g. energy consumed on computational components and on transmission lines, the discharging characteristics of batteries, etc) have a significant impact on the performance, *et_sim* models them accurately with their actual values (see Sec 5 for details).

### 7.1 EAR vs. SDR

Throughout Sec 7.1 to 7.2, it is assumed that a single central controller with infinite energy resource is deployed, and therefore the system dies when some critical node is dead. We will discuss how several central controllers with finite energy affect the system lifetime in Sec 7.3.

In this first set of experiments, a new job is launched when the previous one is completed. In other words, there is exactly one job in the target system and therefore no buffering at nodes is needed. As in Sec 5.1.3, the battery model is based on the discharging characteristics of a thin-film battery together with a discrete-time approximation. The performance data is collected when the target system becomes dead.

The experimental results of performance gain are shown in Fig 7. It is observed that, EAR does better than SDR by a factor between 5 to 15 times, depending on the size of the network. Compared with the significant performance gain, the overhead caused by the control mechanism is very small: indeed, the percentage of energy consumed on exchanging the control information divided by the total energy consumption in 4x4, 5x5, 6x6, 7x7 and 8x8 mesh networks is only 2.8%, 3.1%, 4.1%, 9.3% and 11.6%, respectively.

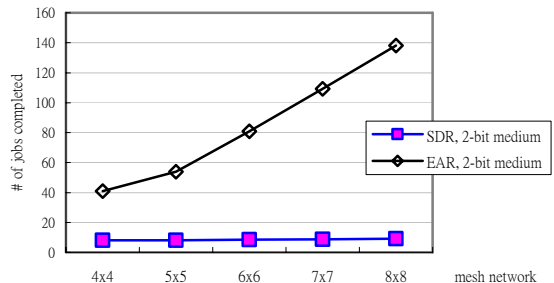

Fig 7: The number of completed jobs (EAR vs. SDR)

### 7.2 EAR vs. Theoretical Upper Bound

In Table 2, the number of jobs completed under EAR are compared with the corresponding analytical upper bounds given by Theorem 1. For a fair comparison, the battery model of the Li-free thin-film battery is replaced with the ideal battery model which outputs constant voltage with 100% efficiency until depletion.

It can be observed in Table 2 that EAR achieves 44.5% to 48.2% of the maximum achievable number of completed jobs. The gap between EAR and the upper bound is due to *a)* the imperfect matching of the mesh topology and application flow, and *b)* the overhead caused by exchanging control information. Therefore, for all practical purposes, EAR performs much better than it appears from data in Table 2.

**Table 2: Comparison between EAR and the upper bound**

| | of jobs completed | Simulation results, $J^{(EAR)}$ | Theoretical upper bound, $J^*$ | $J^{(EAR)}/J^*$ |
|---|---|---|---|---|
| Network size | 4x4 | 62.8 | 131.42 | 47.8% |
| | 5x5 | 92 | 205.25 | 44.8% |
| | 6x6 | 132.7 | 295.70 | 44.9% |
| | 7x7 | 194 | 402.48 | 48.2% |
| | 8x8 | 234 | 525.69 | 44.5% |

### 7.3 Effect of Controller Failures on System Lifetime

In this section, the effect of controller failures due to battery depletion is investigated. We designed in Verilog several central controllers for mesh networks of various sizes and measured their power consumption. For example, the controller of a 4x4 mesh network operating at 100MHz clock frequency consumes a dynamic power of 6.94mW and a leakage power of 0.57mW.

Plotted in Fig 8 are the simulation results for a different number of controllers, each controller having an attached battery as mentioned in Sec 5.1.3. The system lifetime is determined by either the lifetime of the AES nodes or the lifetime of the central controllers, whichever is smaller. For a fixed-sized mesh network, increasing of the number of controllers extends the system lifetime up to a threshold after which the lifetime of AES nodes becomes the dominant factor of the system lifetime. Given a fixed number of available controllers, the tails in Fig 8 are decreasing because a controller for a bigger mesh consumes more power than a controller for a smaller mesh.

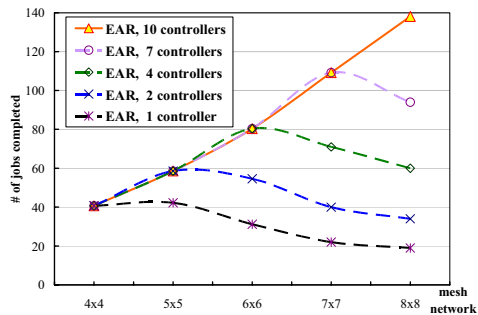

Fig 8: The effect of the amount of central controllers on system lifetime in mesh networks of various sizes

### 8. Conclusion

In this paper, the issue of energy-aware routing in electronic textiles has been addressed. As main contributions, we have derived an analytical upper bound for routing and developed a general-purpose energy-aware routing algorithm, EAR, which considers the concrete limitations given by limited battery capacities, long wire effects, etc. For the AES cipher, EAR achieves approximately 50% of the analytical upper bound. Furthermore, in a mesh network with less or equal to 64 nodes, the performance gain of EAR over its non-energy-aware counterpart ranges from 5 to 15 times, depending on the size of the network. It is expected that the performance gain will be even higher in larger networks.

### References

[1] E. R. Post and M. Orth, "Smart Fabric, or Wearable Clothing," Proc. of Int'l Symp. on Wearable Computers, Oct. 1997.
[2] E. R. Post, M. Orth, P.R. Russo, and N. Gershenfeld, "E-Broidery: Design and Fabrication of Textile-Based Computing," IBM Systems J., Vol. 39, nos. 3/4, pp. 840-860, 2000.
[3] M. Gorlick, "Electric Suspenders: A Fabric Power Bus and Data Network for Wearable Digital Devices," Proc. Third Int'l Symp. Wearable Computers, Oct. 1999.
[4] D. Marculescu, et al., "Electronic Textiles: A Platform for Pervasive Computing," in Proc. of the IEEE, Dec. 2003.
[5] P. Stanley-Marbell, et al., "Modeling, Analysis and Self-Management of Electronic Textiles," in IEEE Trans. on Computers, Vol. 52, Issue 8, Aug. 2003.
[6] D. Cottet, J. Grzyb, T. Kirstein, and G. Tröster, "Electrical Characterization of Textile Transmission Line," IEEE Trans. on Advanced Packaging, Vol. 26, Issue 2, May 2003.
[7] FIPS 197, announced by NIST, Nov. 2001
[8] L. Benini, G. Castelli, A. Macii, E. Macii, M. Poncino, and R. Scarsi, "Discrete-Time Battery Models for System-Level Low-Power Design," IEEE Trans. on VLSI Systems, Oct. 2001.
[9] T. H. Cormen, C.E. Leiserson, and R. L. Rivest, "Introduction to Algorithms," McGraw-Hill, 1990.
[10] B. J. Neudecker, N. J. Dudney, and J. B. Bates, "Lithium-Free Thin-Film Battery with In Situ Plated Li Anode," Journal of the Electrochemical Society, 147 (2) 517-523.
[11] B. J. Neudecker, M. H. Benson and B. K. Emerson, "Power Fibers: Thin-Film Batteries on Fiber Substrates," http://www.darpa.mil/dso/thrust/matdev/smfm/Present.html
[12] M. Maleki, K. Dantu, and M. Pedram, "Lifetime Prediction Routing in Mobile Ad Hoc Networks," Proc. of IEEE Wireless Communication and Networking Conf., Mar. 2003.
[13] J.-H. Chang and L. Tassiulas, "Maximum Lifetime Routing in Wireless Sensor Networks," in IEEE/ACM Trans on Networking, Vol. 12, No. 4, Aug 2004.